\documentclass[reprint,amsmath,amssymb,aps,pra,floatfix]{revtex4-1}

\usepackage{graphicx}
\usepackage{dcolumn}
\usepackage{bm}
\usepackage{tabularx}
\begin{document}

\title{The Static Dipole Polarizability of Palladium from Relativistic Coupled Cluster Theory}

\author{Paul Jerabek}\email{paul.jerabek@gmail.com}
\affiliation{Centre for Theoretical Chemistry and Physics,
 The New Zealand Institute for Advanced Study and the Institute for Natural and Mathematical Sciences \\
 Massey University Albany, Auckland (New Zealand)}
\author{Peter Schwerdtfeger}\email{p.a.schwerdtfeger@massey.ac.nz}
\affiliation{Centre for Theoretical Chemistry and Physics,
 The New Zealand Institute for Advanced Study and the Institute for Natural and Mathematical Sciences \\
 Massey University Albany, Auckland (New Zealand)}
 \affiliation{Centre for Advanced Study (CAS) at the Norwegian Academy of Science and Letters, Drammensveien 78, NO-0271 Oslo, Norway}
 \author{Jeffrey K. Nagle}\email{jnagle@bowdoin.edu}
 \affiliation{Department of Chemistry, Bowdoin College, 6600 College Station Brunswick, Maine 04011 (USA)}

\date{\today}

\begin{abstract}
Nonrelativistic and relativistic coupled-cluster calculations extrapolated to the complete basis set limit including excitations up to the quintuple level (CCSDTQP) were carried out to accurately determine the static electric dipole polarizability of the closed-shell palladium atom. The resulting value of $\alpha$ = 26.14(10) a.u. implies that palladium has the smallest dipole polarizability of all known elemental metal atoms due to its unique 4d$^{10}$5s$^0$ configuration. Relativistic effects are found to be already sizeable ($\Delta_R\alpha$= +1.86~a.u.) compared to electron correlation ($\Delta_C\alpha$= +5.06~a.u.), and need to be included for the accurate determination of the dipole polarizability. We also report a value of the second hyperpolarizability to be $\gamma\approx$ 40,000~a.u., but here the coupled-cluster contributions are not yet converged out with respect to higher than quintuple excitations.
\end{abstract}

\pacs{31.15.ap,31.15.A-,31.30.jc,31.15.bw}
\maketitle

\section{Introduction}

The electric dipole polarizability $\alpha$ is one of the most fundamental atomic properties important for many atomic and molecular properties and applications \cite{bonin1997,Bishop1999,maroulis2004,maroulis2006,Mitroy2010,champagne2010,Safronova2015}. Whereas we have some knowledge of static dipole polarizabilities of all the known elements in the Periodic Table up to high nuclear charge, its accurate determination still remains a considerable challenge for both experiment and theory \cite{Hohm-2000,schwerdtfeger2006,Thierfelder2009,Mitroy2010}. This is especially the case when open-shell atoms are considered where, beside the scalar, the tensor component in the correct coupling scheme needs to be taken into account \cite{Fleig-2005,Thierfelder2008,Safronova2015}. On the other hand, considerable progress has been made over the past two decades in the accurate determination of closed-shell static dipole polarizabilities of the neutral Group 2, 12 and 18 elements of the Periodic Table, both from theory and experiment \cite{Safronova2015,schwerdtfeger2017table}.

Palladium is a rare element that is used in many applications such as catalytic converters and fuel cell technologies. The valence ground state electron configuration of atomic palladium is closed-shell 4$d^{10}$5$s^0$, differing from all the other Group 10 members Ni (3$d^8$5$s^2$), Pt (5$d^9$6$s^1$), and Ds (6d$^8$7$s^2$), which are open-shell \cite{Dzuba2016a,Demissie2018a}. In fact, Pd is the only known atom in its ground electronic state not to have at least one electron in an outer-shell n$s$ or n$p$ orbital \cite{Kramida}. Given this unique feature, it is particularly important to have reliable values of its fundamental properties such as ionization energy, electron affinity, and static electric dipole polarizability. While accurate experimental values for both its ionization energy \cite{Kramida} and electron affinity \cite{Scheer1998} are available, we are not aware of any experimental determination of its dipole polarizability. Recently published theoretical and empirical estimates differ widely from about 13--62~a.u., see Table \ref{tab:pd-polarizability-overview}. If the polarizability is actually less than about 30~a.u. it would make palladium, along with the superheavy elements copernicium ($Z=112$, $\alpha$=27.64 a.u. \cite{Pershina2008}) or nihonium ($Z=113$, $\alpha$=29.9 a.u. \cite{Pershina2008b}), a contender for having the smallest polarizability of any metal atom in the periodic table. Moreover, an accurate value of the dipole polarizability helps to benchmark other methods such as density functional theory \cite{Bast2008b}.

For an accurate quantum theoretical treatment of the electronic response to an applied external electric field, both relativistic and electron correlation have to be taken into account \cite{Schwerdtfeger1994,Lim1999,Thierfelder2008,Thierfelder2009}. For palladium, the main contribution from the sum-over-states formula of the dipole polarizability will come from 4d$\rightarrow$5p excitations, and we may therefore expect relativistic effects to be rather small but not negligible for an accurate electronic structure treatment. The present calculations were undertaken in an attempt to establish an accurate value for the dipole polarizability of the closed-shell palladium atom using relativistic coupled cluster theory. We also provide a value of the fourth-order term with respect to the applied electric field, the (second) hyperpolarizability $\gamma$, although higher order derivatives with respect to the electric field are known to be more problematic from a computational point of view \cite{Kassimi1994}.

\begin{table}
 \centering
 \caption{Reported literature values for the static electric dipole polarizability $\alpha$ of Pd (all values in atomic units). Abbreviations used: NR: Nonrelativistic, R: relativistic, DKH2: relativistic effects from second-order Douglas-Kroll-Hess, Dirac: Dirac-Coulomb Hamiltonian, SR-ECP: scalar-relativistic effective-core potential, HF: Hartree-Fock, MP2: second-order M{\o}ller-Plesset, CCSD: coupled cluster with single and double excitations, RPA: random phase approximation, TD-DFT: time-dependent density functional theory, LDA: local density approximation, CAMB3LYP: Coulomb attenuated B3LYP functional, PGG: Petersilka-Gossmann-Gross kernel \cite{Petersilka-1996}, IP: ionization potential.}
 \label{tab:pd-polarizability-overview}
 \begin{tabular}{lll}
  \hline
      $\alpha$ (a.u.)   &  Comments                             & Refs.                    \\
  \hline
{\it ab-initio} \\
      23.1             	& NR-HF                           		& \cite{Fraga1973}         \\
      75.6          	& NR-HF ($^3F$ state, $d^8s^2$)      	& \cite{Thorhallsson1968}  \\
      21.15            	& R-Dirac-HF                            & \cite{Bast2008b}         \\
      21.17      		& R-RPA                               	& \cite{Johnson1983}       \\
      24.581           	& R-DKH2-MP2                			& \cite{Granatier2011a}    \\  
      26.612    		& SR-ECP-CCSD                   		& \cite{Bast2008b}         \\
{\it DFT} \\   
      32          		& R-Dirac-LDA                     	  	& \cite{Miller2002,Doolen-1987}        \\
      30.15         	& R-Dirac-LDA                     	  	& \cite{Bast2008b}         \\
      26.60         	& R-Dirac-CAMB3LYP                     	& \cite{Bast2008b}         \\
      20.94--31.61		& R-Dirac, various DFT approx.          & \cite{Bast2008b}         \\
      61.7         		& TD-DFT(PGG)                          	& \cite{Gould2016b}        \\
      20.0         		& TD-DFT(PGG)                      		& \cite{Gould2016c}        \\
{\it empirical} \\
      32$\pm$6       	& Empirical, IP correlation        		& \cite{Fricke1986}        \\
      58.8        		& Empirical, Slater rules             	& \cite{Ghosh2006a}        \\
      12.84        		& Empirical, IP + radius correlations   & \cite{Hohm2012}          \\
      47$\pm$24     	& NR-HF, scaled                   		& \cite{Miller1978}        \\
  \hline
  \hline
 \end{tabular}
\end{table}

\section{Computational Method}

The total energy in an homogeneous electric field for a closed-shell atom $E(F)$ can be written as (electric field set arbitrarily in $z$-direction, $F=F_z$),
\begin{equation}
\label{eq:e-of-f-for-atoms}
 E(F) = E(0) + \frac{1}{2} \frac{\partial^2 E(F)}{\partial^2 F}\bigg\rvert_{0}F^2  + 
 \frac{1}{24} \frac{\partial^4 E(F)}{\partial^4 F}\bigg\rvert_{0}F^4 \cdots \, \,.
\end{equation}
with the static electric dipole polarizability $\alpha$ and (second) hyperpolarizability $\gamma$ defined as
\begin{equation}
\label{eq:polarizability-in-finite-field}
 \alpha = -\frac{\partial^2 E(F)}{\partial^2 F}\bigg\rvert_{F=0} \quad {\rm and} \quad \gamma = -\frac{\partial^4 E(F)}{\partial^4 F}\bigg\rvert_{F=0} \, \,.
\end{equation}
We computed the electronic energies of Pd in external electric fields (see section above) with field strengths in the range $F=[0.0,0.002]$ a.u. and step size $\Delta F=0.00025$ a.u. at increasing levels of theory to get insight into how much the inclusion of relativistic and correlation effects, and specifically higher electronic excitations, influence the static atomic dipole polarizability. We used relativistic coupled-cluster theory which included excitations from singles, doubles and perturbative triples (CCSD(T)) as implemented in DIRAC-15 \cite{Dirac15,Visscher2001}. All 46 electrons and virtual orbitals up to 30~a.u. were considered in the correlation treatment. Calculations were performed with doubly-augmented, uncontracted, all-electron, triple- and quadruple-$\zeta$ (TZ and QZ) quality basis sets dyall.ae3z [30$s$22$p$15$d$8$f$5$g$] and dyall.ae4z [35$s$27$p$19$d$12$f$8$g$5$h$], respectively  \cite{Dyall2007,Dyall2012}. The energies were then extrapolated to the complete basis set (CBS) limit using two-point extrapolation schemes utilizing exponential extrapolation for the Hartree-Fock energies \cite{Zhong2008} and inverse cubic extrapolation for the correlation energies \cite{Helgaker1997a}, respectively. We used values of 5.79 and 3.05 for $\alpha_{34}$ and $\beta_{34}$, respectively, as proposed by Neese \textit{et. al}. \cite{Neese2011}.

The CCSD(T)/CBS calculations were performed non-relativistically, as well as with inclusion of scalar-relativistic effects (X2C-Spinfree) \cite{Saue2011} and two-component, which includes spin-orbit coupling, and are denoted CCSD(T)\textsubscript{NR}, CCSD(T)\textsubscript{SR} and CCSD(T)\textsubscript{SO}, respectively. The two-component calculations were carried out using the exact two-component X2C-mmf+Gaunt Hamiltonian of DIRAC-15, obtained from a transformation to a two-spinor basis after solving the four-component Dirac-Hartree-Fock equations.\cite{Sikkema2009}

Higher order excitations of the valence electrons were calculated as correction terms to the atomic energies at the scalar-relativistic level of theory, using the second-order Douglas-Kroll-Hess (DKH2) Hamiltonian \cite{Douglas1974,Hess1986,Jansen1989,Reiher2004,Reiher2004a,Peng2012}, and subsequently added to the CBS limit CCSD(T)\textsubscript{X2C} energies. Here we utilized Molpro 2015.1 \cite{Werner,Werner2012,Hampel1992,Deegan1994} in conjunction with the multi-reference coupled cluster code MRCC \cite{Kallay,Rolik2013,Kallay2005a,Kallay2008,Bomble2005}. An augmented, correlation-consistent, core-valence, Douglas-Kroll-Hess basis set aug-cc-pwCVTZ-DK \cite{Peterson2007} was used for these calculations. While we could correlate all electrons in the coupled-cluster calculations with full triples (CCSDT), we had to restrict the active occupied space to the 4$d$ electrons at the full quintuples level of theory (CCSDTQP).

To obtain the individual correction terms, we subtracted the lower-level result from the higher-level one. The full triple correction $\Delta$[T -- (T)]\textsubscript{SR} was obtained by subtracting the perturbative triple CCSD(T)-DKH2\textsubscript{(AE)}/aug-cc-pwCVTZ-DK contribution from full triple energy CCSDT-DKH2\textsubscript{(AE)}/aug-cc-pwCVTZ-DK. For the full quadruple corrections $\Delta$[Q -- T]\textsubscript{SR} we took the energy difference between the CCSDTQ-DKH2\textsubscript{(4d)}/aug-cc-pwCVTZ-DK and CCSDT-DKH2\textsubscript{(4d)}/aug-cc-pwCVTZ-DK calculations with only the $4d$ electrons correlated. The same procedure was applied for the Quintuples correction. However, the CCSDTQ(P) calculations with the TZ basis set were already at the limit of our computational resources, and for the full quintuples we had to restrict the basis set to double-$\zeta$ (DZ) quality. 

\section{Results and Discussion}

The results of our calculations are shown in Table \ref{tab:pd-polarizability}. Scalar-relativistic effects lead to a non-negligible increase in $\alpha$ of 1.482 a.u. at the HF level and 1.836~a.u. at the CCSD(T) level of theory, which originate from the relativistic $4d$ orbital expansion. Spin-orbit coupling increases the dipole polarizability only by 0.032 and 0.026~a.u. at the HF and CCSD(T) level of theory, respectively. Electron correlation contributes 5.056~a.u. to $\alpha$ at the relativistic level. Out of this, 2.067 a.u. come from perturbative triples, 0.101~a.u. from the variational contribution to the triple correction (correcting the  perturbative treatment of triples in CCSD(T)), while the quadruple and quintuple corrections are responsible for raising the polarizability of Pd by 0.079~a.u.
Our final value of 26.135~a.u. for the atomic polarizability of Pd is in good agreement with the, for example, recently reported non-relativistic, effective core potential CCSD value of 26.612~a.u, \cite{Bast2008b} the DK-MP2 relativistic value of 24.581~a.u. \cite{Granatier2011a} and various DFT-calculated values (e.g., the value of 26.60~a.u. from a CAMB3LYP DFT calculations \cite{Bast2008b}, see Table \ref{tab:pd-polarizability-overview}). 
\begin{table}[htb!]
 \centering
 \caption{Nonrelativistic (NR), scalar relativistic (SR) and X2C/Gaunt relativistic (R) atomic polarizabilities $\alpha$ and hyperpolarizabilities $\gamma$ (in atomic units) of Pd computed with the finite field method at different levels of theory.\footnote{Terminology is as follows: CCSD(T)\textsubscript{NR} = CCSD(T)\textsubscript{(AE)}/CBS; CCSD(T)\textsubscript{SR} = CCSD(T)-X2C-Spinfree\textsubscript{(AE)}/CBS, CCSD(T)\textsubscript{X2C} = CCSD(T)-X2C-Gaunt\textsubscript{(AE)}/CBS; $\Delta$[T -- (T)]\textsubscript{SR} = CCSDT-DKH2\textsubscript{(AE)}/TZ -- CCSD(T)-DKH2\textsubscript{(AE)}/TZ; $\Delta$[Q -- T]\textsubscript{SR} = CCSDTQ-DKH2\textsubscript{(4d)}/TZ -- CCSDT-DKH2\textsubscript{(4d)}/TZ; $\Delta$[P -- Q]\textsubscript{SR} = CCSDTQ(P)-DKH2\textsubscript{(4d)}/TZ -- CCSDTQ-DKH2\textsubscript{(4d)}/TZ + CCSDTQP-DKH2\textsubscript{(4d)}/DZ -- CCSDTQ(P)-DKH2\textsubscript{(4d)}/DZ.}}
 \label{tab:pd-polarizability}
 \begin{tabular}{lccc|ccc}
  \hline
  		& \multicolumn{3}{c|}{$\alpha$} & \multicolumn{3}{c} {$\gamma$} \\
   		& NR & SR & R & NR & SR & R\\
  \hline
  HF TZ			&	19.612	&	21.109	&	21.142	&	19252	& 	13381	&	14203	\\
  HF QZ			&	19.575	&	21.060	&	21.092	&	19311	& 	12527	&	14591	\\
  HF CBS		&	19.565	&	21.047	&	21.079	&	20149	& 	12328	&	14669	\\
  CCSD TZ		&	22.898	&	24.560	&	24.587	&	27364	& 	23855	&	22108	\\
  CCSD QZ		&	22.511	&	24.141	&	24.165	&	27339	& 	21418	&	21002	\\
  CCSD CBS		&	22.252	&	23.864	&	23.888	&	26992	& 	20135	&	19217	\\
  CCSD(T) TZ	&	24.718	&	26.606	&	26.635	&	33071	& 	31640	&	27525	\\
  CCSD(T) QZ	&	24.344	&	26.198	&	26.225	&	34091	& 	30245	&	27941	\\
  CCSD(T) CBS	&	24.093	&	25.929	&	25.955	&	34324	& 	29734	&	28137	\\
  \hline
  $\Delta$[T -- (T)]\textsubscript{SR}	&	&	&	+0.101	&	&	& +3013	\\
  $\Delta$[Q -- T]\textsubscript{SR}	&	&	&	+0.161	&	&	& +5088  \\
  $\Delta$[P -- Q]\textsubscript{SR}	&	&	&	-0.082	&	&	& +3499  \\
  \hline
	Final value	&			&			&	\textbf{26.135}	&	&	& \textbf{39737}	\\
  \hline
  \hline
 \end{tabular}
\end{table}

The atomic properties of all known Group 10 elements are summarized in Table \ref{tab:atomic-data-group-10}. We estimated the uncertainty in our calculations from the size of the quadruples plus quintuples corrections 
(0.079 a.u.) and the error estimated from the finite field and CBS limit extrapolation (-0.018 a.u.) using different number of finite field values. Thus, we estimate the total uncertainty to be 0.10 a.u. We note that the Gaunt term of the Breit operator included in our calculations decreases the polarizability by only 0.05 a.u. at the CCSD(T) level of theory using the augmented quadruple-zeta basis set. Furthermore, we estimated QED contributions using Pyykk\"o and Zhao's local approximation for the self-energy contribution \cite{Pyykko-Zhao-2003}. As one expects from polarizing a fully occupied 4d shell by an external electric field \cite{ThiSch10}, the change in the polarizability is very small (-0.007 eV at the Hartree-Fock level using Dyall's augmented QZ basis set) and well within our given uncertainty.

It is apparent from Table \ref{tab:atomic-data-group-10} that the trend in polarizability values for the four Group 10 elements alternates in magnitude from top (Ni) to bottom (Ds) in group 10 of the periodic table. This trend is not consistent with the steady increase in the values of the ionization energy for these four elements, casting doubt on the practice of using ionization energies to predict polarizabilities. For Pd, the small polarizability value clearly results from the different electron occupation compared to the other elements. On the other hand, the small value of the Ds ($6d^87s^2$) polarizability originates from the strong relativistic $7s$ shell contraction.
\begin{table}[htb!]
\centering
 \caption{Atomic data (ionization potential, IP, electron affinity, EA, and dipole polarizability, $\alpha$) for the Group 10 elements Ni, Pd, Pt and Ds.}
 \label{tab:atomic-data-group-10}
 \begin{tabular}{l|rrr}
  \hline
\hspace{0.8cm}     & \hspace{0.5cm} IP/eV\footnote{Ref. \cite{Kramida}} & \hspace{0.8cm} EA/eV\footnote{Ref. \cite{Scheer1998,Bilodeau1999}.} & \hspace{1.0cm} $\alpha$/a.u. \\
  \hline
Ni 	& 7.639877 & 1.15716 & 49$\pm$3\footnote{Average of recent theoretical values for scalar $\alpha$~ \cite{Chandler1987,Pou-Amerigo,Klos2005}. The uncertainty shows the range of reported values.} \\
Pd 	& 8.33686  & 0.56214 & 26.14(10)\footnote{This work.}   \\
Pt 	& 8.95883  & 2.12510 & 48$\pm$4\footnote{Average of empirical and theoretical values for scalar $\alpha$~
 \cite{Miller2002,Hohm2012,Gould2016b}. The uncertainty shows the range of reported values.} \\
Ds 	& 11.2(1)  & -- & 32(3)\footnote{From Dirac-Hartree-Fock +RPA calculations with use of fractional occupation
numbers \cite{Dzuba2016a}} \\
  \hline
  \hline
 \end{tabular}
\end{table}

We also provide information for the second hyperpolarizability of Pd, see Table \ref{tab:pd-polarizability}. Here we see extremely large relativistic effects due to the indirect coupling between the $4d$ and the relativistically contracted $5s$ orbitals (note that from the sum-over-states formula for the hyperpolarizabilities \cite{Saue2018book} we couple states of angular momentum $l$ with states ranging from $l-2$ to $l+2$ with $l\ge 0$). Electron correlation effects are therefore extremely large, which is well known for hyperpolarizabilities in general \cite{Kassimi1994}. For Pd, the triple, quadruple  and quintuple contributions are so large that an accurate prediction of the hyperpolarizability cannot be made at this stage. Indeed, for such properties sum-over-states Monte Carlo configuration interaction (CI) equivalent to full CI have recently used to obtain close to exact values \cite{Coe-2014}. However, for transition elements this would be a formidable task. Future experimental measurements would therefore be welcome.

\section{Summary}
Our calculated value for the palladium atomic polarizability of 26.14(10)~a.u. is the most accurate obtained so far, and is exceptionally small compared to all other $d$-block and $f$-block atoms. This is a result of its unique 4$d^{10}$5$s^0$ valence electron configuration with a rather compact closed $4d$-shell. The hyperpolarizability is extremely sensitive to both relativistic and electron correlation effects and requires further detailed investigation at even higher correlated level.

\begin{acknowledgements}
We acknowledge financial support by the Alexander-von-Humboldt Foundation (Bonn, Germany) and the
Marsden Fund of the Royal Society of New Zealand (17-MAU-021). JN is grateful to the Bowdoin College for sabbatical leave support. We thank Prof. Trond Saue (Toulouse) for useful discussions.
\end{acknowledgements}

\end{document}